# Confining Windows Inter-Process Communications for OS-Level Virtual Machine

Zhiyong Shan

Stony Brook University, Renmin University of China

zyshan@hotmail.com

Yang Yu   Tzi-cker Chiueh

Stony Brook University

{yyu, chiueh}@cs.sunysb.edu

**ABSTRACT**

As OS-level virtualization technology usually imposes little overhead on virtual machine start-up and running, it provides an excellent choice for building intrusion/fault tolerant applications that require redundancy and frequent invocation. When developing Windows OS-level virtual machine, however, people will inevitably face the challenge of confining Windows Inter-Process Communications (IPC). As IPC on Windows platform is more complex than UNIX style OS and most of the programs on Windows are not open-source, it is difficult to discover all of the performed IPCs and confine them. In this paper, we propose three general principles to confine IPC on Windows OS and a novel IPC confinement mechanism based on the principles. With the mechanism, for the first time from the literature, we successfully virtualized RPC System Service (RPCSS) and Internet Information Server (IIS) on Feather-weight Virtual Machine (FVM). Experimental results demonstrate that multiple IIS web server instances can simultaneously run on single Windows OS with much less performance overhead than other popular VM technology, offering a good basis for constructing dependable system.

**Categories and Subject Descriptors** D.4.5 [**Operating Systems**]:Reliability; D.4.6 [**Operating Systems**]: Security and Protection

**General Terms** Reliability, Security

## 1. Introduction

OS-level virtualization technology usually imposes little or no overhead on virtual machine (VM) start-up, running and shut-down. Therefore, OS-level VM provides an excellent platform for intrusion/fault tolerant applications that require redundancy and frequent invocation. An OS-level VM is able to share as many resources as they can with other VMs and host environment. Meanwhile, programs in VM run as normal applications which directly use the normal system call interface of the host operating system and do not need to be subject to an intermediate virtual machine, as is the case with whole-system virtualizers (e.g., VMware[1]) or paravirtualizers (e.g, Xen[2]).

When developing OS-level virtual machine on the popular Windows OS, in order to achieve strong isolation, the inter-process communications often need to be constrained among processes in the same VM. However, simply confining all IPCs will obviously disturb most processes' running as processes in a VM usually need to cooperate with other processes in the host, especially some system service processes, e.g. Service Control Manager (SCM). These service processes must run in the host environment rather than initiate another instance in VM, because they have too tight relationship with Windows OS itself. Therefore, a proper IPC confinement mechanism is desired to confine IPC within a VM's scope while not disturbing a process' running.

Building such an IPC confinement mechanism is, however, a challenging task. IPC on Windows platform is more complicated than that of UNIX style OS, including mutex, event, timer, semaphore, shared memory, mailslot, pipe, socket, RPC, LPC, DDE, COM, Windows message, data copy, clipboard, etc. Furthermore, as most of the programs on Windows are not open source, it is difficult to discover all of the performed IPCs of a running process. In other words, it is difficult to figure out which IPC should be confined within VM and which IPC should not. Consequently, building IPC confinement mechanism would involve tedious program activities tracing, comparing and analyzing to uncover all of the used IPCs of programs.

As far as we know, there is no proper IPC confinement mechanism for Windows OS in the literature. There are only two projects similar to our work. One is Feather-weight Virtual Machine (FVM) [3] that enables multiple isolated execution environments to run on a single Windows kernel. It can correctly confine IPC for some processes, however, it fails to confine IPC for other processes that have to communicate with system service in host (e.g. RPCSS service process and IIS service processes) and therefore fails to virtualize these important services. The other project is Virtuozzo [5] that provides isolated environments called Virtual Dedicated Server or Virtual Private Server on Windows platform, but we could not find any descriptions about Windows IPC confinement from their public documents.

In this paper, we propose three general principles to confine IPCs on Windows OS and a novel Windows IPC confinement mechanism based on the principles. The proposed confinement mechanism not only can confine IPCs within a VM's scope, but also can correctly facilitate all IPCs between processes in a VM and in host so that executions of the processes in a VM are not disturbed. It employs two tables to help automatically and efficiently identify these IPCs. One table records IPC objects created by system services in host, through which a process in the VM is able to talk with the system services. The other table records the global IPC objects created inside a VM, which should only be accessed by processes in the same VM instead of processes in other VMs or the host. Based on the mechanism, we further successfully virtualized the critical system service RPCSS and popular web server IIS. From literature studies, this is the first time that multiple RPCSS and IIS instances can be successfully run on a single Windows OS, which is usually prevented by Windows OS.

In the rest of the paper, we firstly describe the background of this work in Section 2, then present the IPC confinement principles and enforcement issues, as well as integrated mechanism in Section 3. Application and test are presented at

Section 4, where we successfully virtualize several important Windows services based on the enforcement of IPC confinement mechanism, and three performance experiments show that it only incurs small additional performance overhead. We provide related work in Section 5 and conclude our work in Section 6.

## 2. Background and Challenge

In order to achieve IPC confinement and build a basis for constructing dependable system, we choose FVM [3], a typical Windows OS level virtualization technology, as the fundamental framework. The key design goal of FVM is efficient resource sharing among VMs so as to minimize VM start-up/shut-down cost and scale to a larger number of concurrent VM instances. As a result, FVM can be a good platform supporting intrusion-tolerant applications, for instance, "scalable web site testing" [4] that can isolate the potential malicious side effects of browser attacks from untrusted web sites from the underlying host machine.

The key idea behind FVM is namespace virtualization, which renames system resources through a virtualization layer, called FVM layer, at the OS system call interface. Windows supports numerous types of namespaces for various system resources, such as files, registries, kernel objects, network address, Windows services, window classes, etc. The FVM layer manipulates the names of all these resources when a process makes system calls to access them. Through resource renaming, the namespaces visible to processes in one VM are guaranteed to be disjoint from those visible to processes in another VM. As a result, two VMs never share any resources and therefore cannot interact with each other directly. For example, suppose an application in one VM (say vm1) tries to access a file /a/b, then the FVM layer will redirect it to access /vm1/a/b. When a process in another VM (say vm2) accesses /a/b, it will try a different file, i.e., /vm2/a/b, which is different from the file /a/b in vm1.

However, completely separating namespaces of different VMs may require unnecessary duplication of common system resources and may lead to the same performance overhead as many heavyweight virtual machine technologies. Being feather-weight, the FVM architecture enables VMs to share most resources with the host environment while isolating state changes of each VM through a special copy-on-write scheme. A newly created VM initially can share all the resources of the host machine. Later on, if processes in the VM make only read requests to system resources, they can simply access the shared resources on the host machine. The VM does not occupy any private resources until processes in the VM try to modify the host machine's resources. Therefore, the resource requirement of each VM is significantly reduced under the FVM architecture.

In the previous version of FVM, however, there were several unresolved issues, such as RPCSS could not run inside VM, IIS web server could not run inside VM, Microsoft office assistant and some installation programs could not work inside VM, etc. The root reason of these issues is that FVM did not have a proper IPC confinement mechanism to not only confine IPC within a VM but also correctly facilitate communications between processes in VMs and in host. As Windows programs' internal details and complex Windows IPC mechanism are generally not documented, building a proper IPC confinement mechanism becomes a challenge in FVM development group.

## 3. Windows IPC Confinement

The Windows operating system provides mechanisms for facilitating communications and data sharing between applications. Collectively, the activities enabled by these mechanisms are called inter-process communications (IPC). Some forms of IPC facilitate the division of labor among several specialized processes. Other forms of IPC facilitate the division of labor among computers on a network.

There are miscellaneous IPC objects in Windows OS. About 18 types of methods can be used for inter-process communication, excluding communications through ordinary file and registry. To facilitate the research, we categorize them into following seven groups according to their internal mechanisms: I. Port related IPC: LPC, RPC, COM/DCOM/COM+; II. Pseudo file related IPC: Mail slot, Pipes; III. Shared memory: File mapping; IV. Synchronization IPC: Semaphore, Mutex, Event, Timer; V. Message related IPC: Windows message, Data Copy, Clipboard, DDE; VI. Windows Sockets: Socket; VII. Dangerous functions: Find Window, Create remote thread, Set window hook.

### 3.1 IPC Confinement Principles

Before designing the IPC confinement mechanism, we provide three general principles to confine Windows OS IPC, based on our work of tracing and analyzing a group of Windows processes' activities.

(1) **Isolation-Principle**: allowing inter-process communications within a VM's scope while blocking the ones across VM borders as much as possible.

To achieve strong isolation, IPC confinement requires that a process running in one VM does not communicate with processes running in other VMs or in the host machine through IPC. This is the basic requirement of IPC confinement. However, only enforcing this principle would result in a failure when a process in VM needs to talk with a process in host. So we need other principles to facilitate necessary process communications across VM border.

(2) **Global-Object-Principle**: allowing processes in a VM to access any global-object except that the global-object is created by a process in different VM.

Global-object refers to the IPC object that can be shared by all processes on the single OS. Conventionally, the global-objects have to be created by users with a special right. In most cases, system services create the global-objects, through which system services provide functionality to all applications on the OS. So, applications in all VMs should also be able to access the global-object by default.

However, once a system service itself is virtualized, i.e. a new instance of the system service is running in VM, or a process running in VM with the special right to create global-object, the global-objects created by these processes should not be accessed by processes in other VMs or host, in order to achieve strong isolation. This is because a virtualized system service running inside a VM should only serve for applications within the same VM, and a process with the special right also should only communicates with other processes within the same VM.

(3) **Host-Object-Principle**: allowing processes in VM to access IPC objects created by a system service in host.

Many applications running in VM need to co-operate with system services in host so that they can proceed with their executions. In most cases, system services utilize non-global IPC objects to provide functionalities to applications rather than global-objects mentioned above. For these non-global IPC objects, called host-objects, we should allow processes in VM to access them.

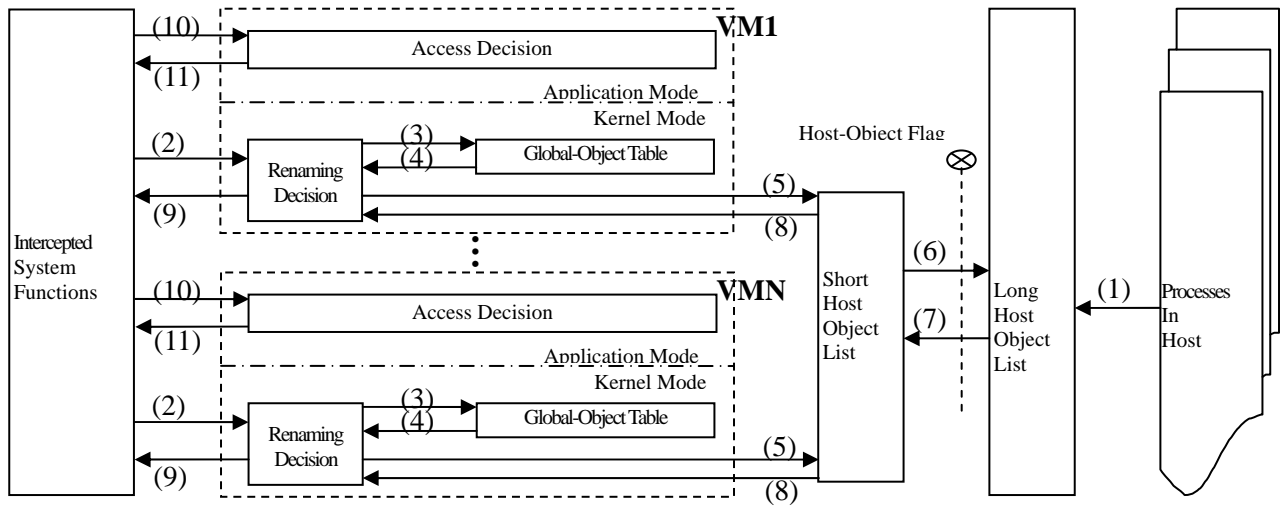

**Figure 1. IPC Confinement Mechanism**

### 3.2 Issues to Enforce the Principles

**How to enforce Isolation-Principle?** The difficulty of enforcing the Isolation-Principle lies in the fact that there are miscellaneous IPC objects with distinct internal mechanisms. For the IPC types I to IV, we intercept related system call functions in kernel level and employ the "rename" method presented in section 2 to enforce Isolation-Principle. Through IPC object renaming, the IPC object visible to processes in one VM are guaranteed to be disjoint from those visible to processes in another VM. As a result, two VMs never share any IPC objects and therefore cannot interact with each other directly.

However, for other IPC types, we intercept related API functions at application level and employ different methods to enforce Isolation-Principle since their mechanisms are different. For IPC type V, i.e. message related IPC, we directly block the message when the sender and the receiver are in different VMs. For IPC type VI, i.e. Windows Socket, we assign an exclusive IP address to a VM by employing the method of IP aliasing and associate the VM's IP address with a socket. For IPC type VII, i.e. dangerous functions, we directly prohibit a process from creating remote thread, modifying other process' address space, setting system wide hooks across VM scope, and enumerating windows in other VMs or host.

**How to enforce Global-Object-Principle?** The difficulty of enforcing the Global-Object-Principle is to correctly and thoroughly identify the global-objects created by virtualized system services or processes with the special right. To address the issue, we construct a table, named Global-Object Table, in each VM to record all global-objects created within the VM. Every time a process in a VM tries to access a global-object, we check if it is in the table. If this is true, we direct the access to the global-object created within the VM. Otherwise, direct the access to the global-object in host.

**How to enforce Host-Object-Principle?** The difficulty of enforcing Host-Object-Principle is how to correctly and thoroughly identify all of the host-objects created by system services in host. After manually tracing and analyzing the activities of processes which have to communicate with system services in host, we successfully identified the host-objects and hard-coded them into FVM source codes. As a result, these processes were able to run inside a VM. But, this method has two limitations. One is that not all of the host-objects can be manually found out, thus the method can not handle all kinds of applications. The other is that the method is platform dependent. Once moving FVM to a new Windows version, we have to look for the host-objects again.

In order to address the issue automatically and efficiently, we design a table, named Host-Object Table, to record all host-objects of the whole system. It consists of two lists. One is a long list to store all the IPC objects of system services in host. However, the long list is usually too long to find out an object in a short time thus impacting system performance significantly. So we further design a short list to store most recently used host-objects in order to reduce the time for finding a host-object. When FVM is booted up, the host-objects are read from system services' process space to the long list. Every time a process in VM requests access to an IPC object, it searches the short list first, trying to find the object. Once it fails, it then searches the long list. If successfully obtaining the object from the long list, it records the object in the short list.

However, when accessing an IPC object that is not a host-object, we have to always search in both the short and long lists to make sure it is not a host-object, which is a time consuming procedure. To avoid searching the long list, we performed a serial of experiments showing that the short list will not be updated any more after finishing startup of all virtualized Windows services. This means that all of the host-objects on the OS are selected from the long list to the short list at this time. Hence we set up a flag, named Host-Object Flag, to stop searching the long list after finishing startup of all virtualized Windows services.

Therefore, with the two host-object lists and the Host-Object Flag, all of the host-objects can be automatically identified and efficiently retrieved by the processes requiring to access them.

### 3.3 IPC Confinement Mechanism

With the IPC confinement principles and their enforcement methods discussed above, the IPC confinement mechanism is carefully designed, as shown in Figure 1.

The IPC confinement mechanism comprises of a set of modules, including Renaming Decision, Access Decision, Global-Object Table, Short Host-Object List, Long Host-Object List and

Host-Object Flag. According to the Isolation-Principle and its enforcement method, the Renaming Decision module is responsible for determining whether renaming IPC objects involving types I to IV. The Access Decision module is responsible for determining whether to allow accessing IPC objects that involves types V to VII. According to the Global-Object-Principle and its enforcement method, the Global-Object Table stores global-object names to help the Renaming Decision module to make renaming decision about global-object. According to the Host-Object-Principle and its enforcement method, the Short Host-Object List, Long Host-Object List and Host-Object Flag help the Renaming Decision module to make renaming decision about host-object.

The bracketed numbers in Figure 1 represent the working steps of the mechanism. In step (1), when FVM is started, all names of the IPC objects except global-objects created by the system service processes in host are read and stored into the Long Host-Object List. In steps (2) and (9), when in kernel mode, a process in VM tries to access an IPC object whose type is one of I to IV. It sends a request to Renaming Decision module and waits for the decision result. In steps (3) and (4), the Renaming Decision module checks whether the object is in the Global-Object Table which stores names of the global-objects created by processes in VM. If this is true, the Renaming decision module returns the renamed global-object name. In steps (5) and (8), Renaming Decision module checks whether the object is in the Short Host-Object List which stores the names of the host-objects recently used and returns the original host-object name if it is true. If the object is not in the Short Host-Object List and the Host-Object Flag is on, the Renaming Decision Module does not search the Long Host-Object List any more and returns the renamed object name. In steps (6) and (7), if it failed to find the object from the Short Host-Object List while the Host-Object Flag is off, Renaming Decision module then searches the Long Host-Object List which stores the names of all IPC objects created by the system services in host. If the object is found, it stores the object name into the short list and returns the original object name; otherwise, it returns renamed object name. In steps (10) and (11), in application mode, a process in VM tries to access an IPC object whose type is one of V to VII. It sends a request to Access Decision module and waits for the decision result. The Access Decision module makes decision on whether to allow the request directly based on the result of analyzing two process' VM ID.

In summary, the mechanism is able to confine all kinds of IPC while efficiently facilitating necessary IPC between processes in VMs and in host. The Renaming Decision module handles the IPCs of types I to IV in kernel mode and the Access Decision module handles the IPCs of types V to VII in application mode. Meanwhile, the Global-Object Table and Host-Object Table facilitate the IPCs between processes in VM and in host. With the help of these two tables, the mechanism is able to automatically detect all IPCs between processes in VM and in host. Furthermore, with the help of the Short Host-Object List and the Host-Object Flag, the mechanism is able to efficiently retrieve host-objects while avoiding time-consuming searching in the Long Host-Object List.

## 4. Application and Test
### 4.1 Application

In the previous version of FVM, there were several unresolved issues, such as RPCSS could not run inside VM, IIS web server could not run inside VM, office assistant and some installation programs could not work inside VM, etc. The root reason of these issues is that FVM did not have a proper IPC confinement mechanism, which can prevent processes in VM from failing caused by abnormal IPC access.

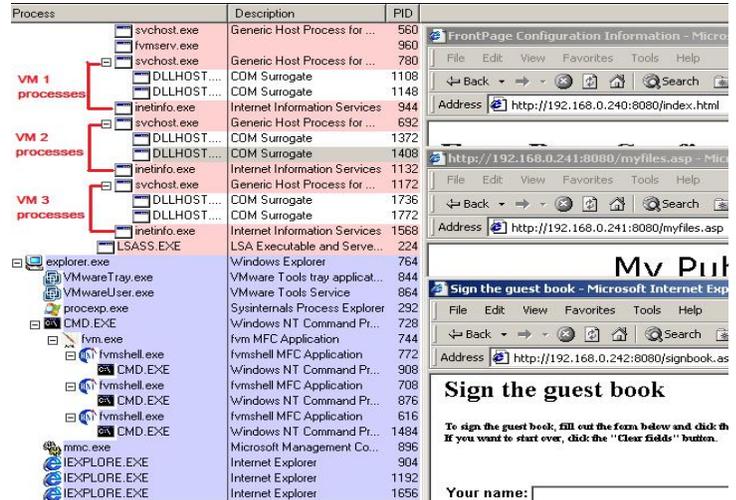

**Figure 2. Three IIS Web Servers Run on Single OS**

With our IPC confinement mechanism enforced, the FVM now can virtualize services including RPCSS on both Windows 2k and XP, IIS web server that contains services IISADMIN and W3SVC on Windows 2k, as well as service Dcomlaunch on Windows XP. Furthermore, a bunch of applications that need these services' support can also run in VM, which proves that these virtualized services act correctly.

Figure 2 shows a snapshot of running three IIS web server instances simultaneously on a single Windows OS. On the left, there is a ProcessExplorer GUI displaying a process list that contains three groups of virtualized IIS services' processes, running in three VMs respectively. In each VM, there is a virtualized RPCSS process svchost.exe, a virtualized IIS web server process inetinfo.exe, and two virtualized DCOM server surrogate processes named DLLHOST.exe that handle ASP web page requests. On the right, there are three IE windows showing both HTML and ASP pages gotten from the three virtualized IIS web servers. As each VM has its own IP address, there is a different IP address displayed in each IE's web address window, which means the IE is accessing an IIS web server running in different VM. In short, this snapshot verifies that IIS web server and RPCSS are successfully virtualized and they work correctly.

To demonstrate the working results of the three IPC confinement principles, Table 1 shows what RPCSS service created IPC objects are confined by the three IPC confinement principles respectively.

**Table 1. IPC objects of RPCSS service**

| Principle | Type | Object |
|---|---|---|
| Isolation | Port | \RPC Control\epmapper<br>\RPC Control\OLE30778CF8A8F24282B5F73ADC0B14 |
| | Named pipe | \Device\NamedPipe\epmapper<br>\Device\NamedPipe\Winsock2\CatalogChangeListener-30c-0 |
| Global-Object | Section | \BaseNamedObjects\Global\RotHintTable |
| Host-Object | Port | \RPC Control\DNSResolver<br>\RPC Control\ntsvcs |
| | Named Pipe | \Device\NamedPipe\net\NtControlPipe* (* represents an arbitrary number)<br>\Device\NamedPipe\svcctl<br>\Device\NamedPipe\ntsvcs<br>\Device\NamedPipe\EVENTLOG |
| Mutex | \BaseNamedObjects\DBWinMutex \BaseNamedObjects\RasPbFile | |
| Section | \BaseNamedObjects\__R_0000000000da_SMem__ \BaseNamedObjects\DBWIN_BUFFER | |
| Event | \BaseNamedObjects\ScmCreatedEvent \SECURITY\LSA_AUTHENTICATION_INITIALIZED | |

**Table 2, IPC Confinement Overhead**

## 4.2 Performance Test

In the following three experiments, we evaluate the overhead of IPC confinement, and the startup overhead of the virtualized services, as well as the performance of redundant IIS web servers. The objective of the last two experiments is to make clear whether the FVM enforced with the IPC confinement mechanism suits for building intrusion/fault tolerant systems which require frequent invocation and redundancy. The test-bed consists of two machines. Machine A is Pentium-4 2.8GHz with 512MB memory running both Windows 2k and XP; machine B is Intel Core 2 Duo CPU 2GHz with 2GB memory running both Windows 2k and XP. Both machines are installed FVM and VMWare Workstation 5.0.

Since the performance overhead of IPC confinement comes

| System calls | Native | FVM | IPC confined FVM | | Difference |
|---|---|---|---|---|---|
| NtOpenSemaphore | 30234 | 64286 | Isolation-Principle | 64493 | 113%, 0.3% |
| | | | Global-Object-Principle | 64388 | 113%, 0.2% |
| | | | Host-Object-Principle | 64471 | 113%, 0.3% |
| NtCreatePort | 37241 | 72309 | Isolation-Principle | 72545 | 95%, 0.3% |
| | | | Global-Object-Principle | 72410 | 94%, 0.1% |
| | | | Host-Object-Principle | 72537 | 95%, 0.3% |
| NtOpenSection | 38134 | 72742 | Isolation-Principle | 72849 | 91%, 0.1% |
| | | | Global-Object-Principle | 72793 | 91%, 0.1% |
| | | | Host-Object-Principle | 72823 | 91%, 0.1% |
| NtCreateNamedPipeFile | 204711 | 269150 | Isolation-Principle | 269535 | 32%, 0.1% |
| | | | Global-Object-Principle | 269251 | 32%, 0.03% |
| | | | Host-Object-Principle | 269323 | 32%, 0.1% |

from the overhead of executing additional instructions associated with every intercepted IPC system calls, we carry out an experiment to measure the overhead of IPC system call interception. We first disable the FVM virtualization layer, run a group of services and applications natively on host environment, and count the average CPU cycles spent in each system call through *rtdsc* instruction. Second, we enable the former version of FVM layer and run the same services and applications in a VM to do the test again. Third, we enable FVM layer which is enforced with the IPC confinement mechanism to perform the test one more time. For the third test, we further take the three IPC confinement principles into account, which is to test three situations corresponding to the three principles for each system call. In both tests, the average CPU cycles of each system call or principle is calculated from 100 invokes. Results are shown in Table 2.

From table 2, we can see that the FVM which is enforced with IPC confinement mechanism takes 32% ~113% more CPU cycles than native. Although the overhead is not small, the impact to the whole system performance is much limited, because the intercepted IPC system calls are merely less than 0.2% of all the invoked system calls according to our program activities tracing and analyzing work. Moreover, compared to the former FVM, the IPC confinement mechanism only adds less than 0.3% extra CPU cycles. Therefore, the general performance impact is small.

The second experiment aims to measure the startup overhead of the virtualized services. Figure 3 shows the startup time of five types of services on machine B, including RPCSS, IISADMIN and W3SVC on 2k, as well as RPCSS and Dcomlaunch on XP. The tests were performed under seven different situations: starting original service in host; starting one, two and three service instances respectively on VMWare; starting one, two and three service instances respectively on FVM. We use a testing program to launch the tested services and record their startup time. The startup time for each service is obtained by measuring the elapsed time from the moment just before calling API OpenSCManager() to the moment when API QueryServiceStatusEx() returns result SERVICE_RUNNING.

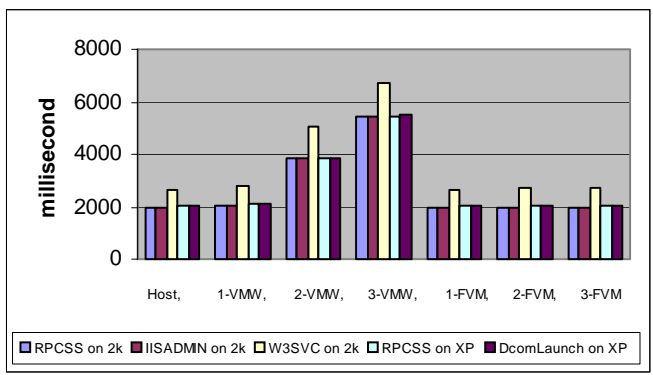

**Figure 3. Services Startup Time**

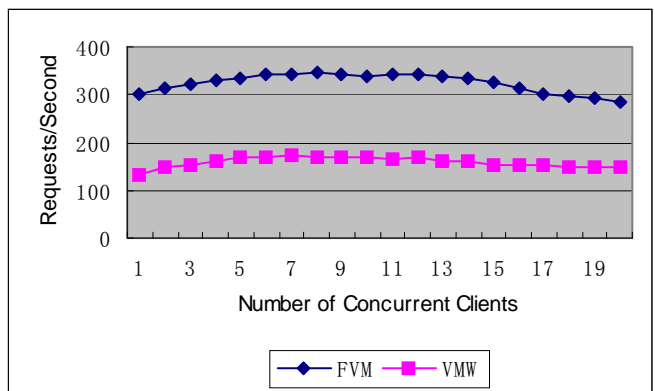

**Figure 4. Performance of Three Running IIS Web Servers**

From Figure 3, the service startup time under the situations of starting one, two and three service instances on FVM are almost equal to that of the original service on host. This indicates that the startup overhead of the virtualized services on FVM is almost zero, and remains very small even when starting multiple virtualized service instances. On the other hand, the service startup time under the situations of starting two and three service instances on VMWare are almost twice or triple that on FVM. This means the overhead of starting two or three service instances on VMWare can be almost twice or triple that on FVM.

The third experiment aims to measure the performance of IIS web servers redundantly running on single OS. Figure 4 shows the performance of IIS web servers under the condition redundantly running three IIS web servers on FVM and VMWare respectively. The test data were collected by Webbench, a licensed PC Magazine benchmark program, from three runs. The IIS web servers and workloads were deployed on machine A, and the configurations of all IIS servers on both FVM and VMWare were the same despite that they may need optimization. On the other side, the Webbench controllers and clients were deployed on machine B. In each testing session, each web server had one to twenty clients concurrently sending requests to it, and each client only had one engine. The testing data are the average of three web servers. The results indicate that the performance of IIS web servers on FVM is as fast as two to three times that of VMWare under the condition running three IIS instances simultaneously on single Windows OS.

In summary, enforcing IPC confinement mechanism imposes small additional performance overhead on Windows OS, while the startup of multiple virtualized services and the performance of IIS web servers redundantly running on single Windows OS are much faster than that of VMWare. Therefore, FVM enforced with IPC confinement mechanism suits better for building intrusion/fault tolerant systems.

## 5. Related work

The former version of FVM was able to confine IPC within VM, but could not correctly handle the inter-process communication between processes in host and in VM. As a result, it failed to run some important services in VM, such as RPCSS and IIS. With our IPC confinement mechanism, those services are successfully virtualized.

There are also other Windows OS level virtualization projects known from literature. SWsoft's Virtuozzo [5] can provide isolated environments called Virtual Dedicated Server or Virtual Private Server on Windows platform, but we could not find any IPC confinement description from their public documents. Some commercial products on Windows with similar virtualization techniques are Softricity Desktop [6] and AppStream [7]. In particular, Softricity Desktop [6] is able to preserve inter-application communications between applications in different VM. Although this can facilitate the applications' executions, it reduces the isolation level. In addition, implementing all IPC confinement at the user-level system library interface makes it easy to be bypassed.

## 6. Summary

This is the first paper to propose the principles and mechanism for Windows IPC confinement which was an unsolved problem in Windows OS-level virtual machine technology. It not only can confine IPC within VM scope, but can also correctly facilitate all necessary IPCs between processes in VM and in host so that executions of the processes in VM are not impacted. It employs two tables to help automatically and efficiently figure out these IPCs. One table records IPC objects created by system services in host, through which processes in VM is able to talk with the system services. The other table records the global IPC objects created inside a VM which should only be accessed by processes in the VM instead of processes in other VMs or in host. Enforced with the IPC confinement mechanism, we can successfully fix the service virtualization problems that previous FVM version could not.

As intrusion/fault tolerant systems require redundancy and frequent invocation, we conducted two experiments on the FVM enforced with IPC confinement mechanism to evaluate the performance of redundant web servers and the overhead of service invocation. Compared to VMWare, the redundant web servers running on FVM are as fast as two to three times that of VMWare. Meanwhile, the service startup overhead is almost zero, and the overhead of starting two or three service instances on FVM can be almost half or one-third of that on VMWare.

In addition, since VMs share most resources with host environment, they can quickly recover from compromised state by discarding the modified resources stored in VM and utilizing resources in host again. Therefore, the properties of being able to run multiple service instances with tiny startup and runtime overhead, as well as being able to quickly recover from fault or intrusion, make the FVM that is enforced with IPC confinement mechanism an excellent building block for dependable systems.

## 7. ACKNOWLEDGMENTS

We thank Prof. Xin Wang at Stony Brook University for her precious advices and the anonymous referees for their useful comments. This research was supported in part by NSF grants CCF-0621512/CNS-0627672, National Science Foundation of China under grant number 60703103/60873213, and the High Technology Foundation of China under grant number 06XNB053.